\documentstyle[preprint,aps]{revtex}

\begin{document}
\draft
\title{Universal spectral correlations at the mobility edge\thanks{To
be published in PRL, vol.\ 72, p.888 (1994)}
}
\author{V. E. Kravtsov$^{1,2}$, I. V. Lerner$^3$, B. L. Altshuler$^4$,
 A. G. Aronov$^{1,5}$ }
\address{$^1$International Centre for Theoretical Physics,
P.O. Box 586, 34100 Trieste, Italy\\
$^2$Institute of Spectroscopy, Russian
 Academy of Sciences, 142092 Troitsk, Moscow r-n, Russia \\
 $^3$School of Physics and Space Research, University of Birmingham,
Birmingham~B15~2TT, United Kingdom\\
$^4$Department of Physics,  Massachusetts Institute of Technology,
 77 Massachusetts Avenue, Cambridge, MA 02139\\
$^5$A.F.Ioffe Physico-Technical Institute,
194021 St.Petersburg, Russia}
\date{20 September 1993}
\maketitle

\begin{abstract}
We demonstrate the level statistics in the vicinity of the Anderson transition
in $d>2$ dimensions to be universal and drastically different from both
Wigner-Dyson in the metallic regime and Poisson in the insulator regime. The
variance of the number of levels $N$ in a given energy interval with $\langle
N\rangle\gg1$ is proved to behave as $\langle N\rangle^\gamma$ where
$\gamma=1-(\nu d)^{-1}$ and $\nu$ is the correlation length exponent. The
inequality $\gamma<1$, shown to be required by an exact sum rule,
results from nontrivial cancellations (due to the causality and scaling
requirements) in calculating the two-level correlation function.
\end{abstract}

\narrowtext

The problem of level statistics in random quantum systems is
  attracting  considerable interest  even
now,  four decades after
the pioneer works of Wigner and Dyson\cite{Wig}.
This is because of the universality of the Wigner-Dyson
statistics which makes it relevant for a large variety of quantum
systems\cite{RMT}.

For the problem of a quantum particle in a random potential, the Wigner-Dyson
statistics is known to be applicable for finite systems in the region
of extended states \cite{GE,Ef:83,AS} which will be referred to as a
metallic region.
With increasing the random potential,  the system undergoes
the Anderson transition into the insulator phase \cite{LR}, where
all states are localized. In this region, the statistics of energy levels is
expected to be  Poisson.

There is, however, the third region, namely,
the critical region in the vicinity of the Anderson transition where
the spectral
statistics is believed to  be still
universal\cite{AS2,ShSh}, although different from both
Wigner-Dyson and Poisson. As the
critical region can not be considered perturbatively or semiclassically,
nearly nothing is known about the third universal statistics.

The first attack at this problem has been done in
 Ref.\cite{AS2} where the simplest
statistical quantity, the variance
$\langle(\delta N)^{2} \rangle$ of the number of energy levels $N$
in a given energy interval of the width $E$, has been considered
($\delta N\equiv N-\langle N\rangle$,
and $\langle \ldots\rangle$ denotes the ensemble average
over the realizations of the random potential).
The dimensional estimation made in Ref.\cite{AS2} has resulted in
$\langle(\delta N)^{2} \rangle=a\langle N\rangle$, thus being different
from the Poisson statistics only by a certain number $a<1$.

We will see, however, that this result contradicts to an exact sum rule
resulting from the conservation of the total number of levels.
The point is that in the dimensional estimations \cite{AS2}
 analytical properties of diffusion propagators have not been
taken into account.
 We will show  that
the analytical properties  resulting from
causality  together with
certain scaling relations near the Anderson transition make
the $ \left< N \right>$--proportional contribution
to the variance  to vanish.
We will calculate the spectral density correlation function and deduce from
it the following {\em universal} relationship between
 the variance and the average number of levels in the energy
interval $E$
\begin{equation}
\label{1}
\left< (\delta N)^{2} \right> =
\frac{a_{d\beta}}{\beta}\langle N\rangle^{\gamma},\qquad \gamma\equiv1-
\frac{1}{\nu d}
\end{equation}
that holds exactly at the mobility edge. Here
$\nu$ is the correlation length exponent,  and
the factor $a_{d\beta}$ is universal in a sense that
it is determined completely by the dimensionality $d$ and
the symmetry
class of the Dyson ensemble ($\beta=1,2,$ or $ 4$ for unitary, orthogonal,
and symplectic ensembles, respectively). For many systems $\nu \approx 1$ so
that $\gamma \approx 2/3$ for $d=3$. In general, Eq.\ (1) suggests a new way
of determining $\nu$.

This is the main result of the paper. It demonstrates that in
the vicinity of the Anderson transition
there really exists
the third universal statistics. It governs the spectral fluctuations that
are weaker than for the Poisson statistics,
$\langle(\delta N)^{2} \rangle_{\rm{P}}\sim\langle N\rangle$, but
much stronger than for the Wigner-Dyson statistics,
$\langle(\delta N)^{2} \rangle_{\rm{W-D}}\sim\ln\langle N\rangle$.

All the three statistics are universal and exact in the same limit:%
 \begin{eqnarray}
 \label{Lm}
 L\rightarrow\infty\qquad E/\Delta=\langle N\rangle={\rm const}\gg 1
 \end{eqnarray}
where $L$ is the sample size. In this limit, the mean level spacing
$\Delta=(\nu_{0}L^{d})^{-1}$ tends to zero
($\nu_{0}$ is the mean density of states), but the number of levels
in an interval $E$ is kept finite, although very large.

The new level statistics describes the fluctuations in an
 energy band $|\varepsilon-\varepsilon_{0}|<E/2$
 centered exactly at the mobility
edge $\varepsilon_{0}\!=\!\varepsilon_c$. For
the critical regime to be achieved
the correlation length $L_{c}(\varepsilon)$ which diverges as
$|{\varepsilon/\varepsilon_{c}-1}|^{-\nu}$ must exceed  the sample size
for {\it all} $\varepsilon$ in the energy band $E$.
Due to this uncertainty $L_c=(E/\varepsilon_{c})^{-\nu}$, and
\begin{equation}
{L_{c}}/{L}=\langle N \rangle^{-\nu}
({L}/{\lambda} )^{d\nu-1},
\end{equation}
 where $\lambda=(\nu_{0}\varepsilon_{c})^{-1/d}$.
Then the Harris criterion \cite{Har}  $\nu>2/d$ ensures
$L_{c}/L\rightarrow \infty$
in the limit (\ref{Lm}) for the energy band centered at
$\varepsilon_{0}=\varepsilon_c$.
In the same limit, the Wigner-Dyson and Poisson statistics describe exactly
the fluctuations in bands centered at
$\varepsilon_{0}>\varepsilon_c$ (the metallic region) and
$\varepsilon_{0}<\varepsilon_c$ (the insulating region), respectively.
The limit  (\ref{Lm}) is required, therefore,
to avoid mixing the levels belonging to different regions as well as to make
the finite-size corrections vanishing.

We consider the  spectral density correlation function
\begin{equation}
\label{R}
R(\omega)\equiv
\frac{1}{\nu_{0}^{2}}
 \Bigl< \nu(\varepsilon)\nu
(\varepsilon+\omega)  \Bigr>-1,
\end{equation}
where $\nu(\varepsilon)$ is the exact density of states at the energy
$\varepsilon$.
 Note that the function
$R(\omega)$ has   a singular term $\delta(\omega)$ resulting from
the self-correlation of  energy levels.

Before deriving the announced result, Eq.\ (\ref{1}), we demonstrate that the
exact sum rule prohibits the variance $\langle (\delta N)^2 \rangle$ to be
$\left< N \right>$--proportional.
The conservation of the
total number of energy levels for any non-singular random
potential may be written down as
\mbox{$\int_{-\infty}^{\infty}[\nu(\varepsilon+\omega)-\nu_{0}] d\omega
=0$}.
It leads to the sum rule:
\begin{equation}
\label{SR}
\int_{-\infty}^{\infty}R(s)\,ds=0,\qquad s\equiv \omega/\Delta.
\end{equation}
The variance $\langle (\delta N)^{2}  \rangle$ of the number of levels
$N$ in the energy band of the width $E$ centered at a certain energy
$\varepsilon_{0}$ (e.g., at the Fermi level $\varepsilon _{{\!}_F}$)
is given by
 \begin{eqnarray}
 \left< (\delta N)^{2} \right>
=\int_{-\left< N\right>}^{\left< N\right>}(\left< N\right>-
|s|)R(s)\, ds.
 \label{V}
 \end{eqnarray}
Then
\begin{equation}
\label{NE}
\frac{d\langle (\delta N)^{2}\rangle}{d\left< N\right>}=
\int_{-\left< N\right>}^{\left< N\right>}R(s)\, ds.
\end{equation}
If the function $R(s)$ is universal in a sense that it does not depend
on any parameter, then the only condition $\left< N\right>\gg 1$ is
sufficient, due to the sum rule  (\ref{SR}),
to make the integral in the r.h.s. of
Eq.\ (\ref{NE}) to be arbitrary small. Therefore, in this case
$\langle (\delta N)^{2}\rangle/\left< N \right>\rightarrow 0$.

The universality assumption is crucial for vanishing the contribution
to the variance proportional to $\left< N \right>$, or the higher power of
$\left< N \right>$. However, a finite disordered sample is characterized by a
set of relevant energy scales that obey in the metallic limit the following
inequalities:%
 \begin{eqnarray}
 \label{neq}
\Delta\ll 1/\tau_D\ll 1/\tau\ll \varepsilon _{{\!}_F}
 \end{eqnarray}
where $\tau_D\!=\!L^2/D$ is the time of
diffusion through the sample, $D$ is the electronic diffusion coefficient in
the classical limit, $D\!=\!v_{{\!}_F}^2\tau/d $, $\tau$ is
the elastic scattering rate, $\hbar\!=\!1$.
Naturally, for sufficiently large $\left< N \right>$ the function $R(s)$
depends not
only on $s$. It results in $\langle (\delta N)^{2}\rangle
\propto(\tau_D\Delta)^{3/2}\left< N \right>^{3/2}$ in
an energy band of the width $E\gg1/\tau_D$\cite{AS}. We will show
elsewhere that higher than $\left< N \right>$ contribution arises also in the
critical region  ($\Delta\!\sim\! 1/\tau_D\ll 1/\tau\!\sim\!
\varepsilon _{{\!}_F}$)
where it is proportional to
$\left< N \right>^{1+\alpha}(\tau\Delta )^{\alpha}$ ($0\!<\!\alpha\!<\!1$
is a certain critical exponent). Both
these nonuniversal contributions could be of importance for finite systems.
However, they do vanish  in the limit (\ref{Lm}).
In this limit only the universal contributions to the
variance survive.

In the insulating regime, the above speculations are not applicable for
estimating the integral in Eq.\ (\ref{NE}). The reason is  the
existence of the additional energy scale $\Delta_{\xi}=1/\nu_{0}\xi^d=
\Delta(L/\xi)^d$
 which is
  a typical spacing for states
confined to a localization volume $\xi^d$ centered at some point.
  Since such states are repelling in the same way as
extended states in metal confined to the whole volume $L^{d}$,
the function $R(s)$ at $s\neq 0$ is
expected to be similar to the Wigner-Dyson function
$R_{W-D}(\omega/\Delta)$
with substituting $\Delta$ by $\Delta_{\xi}$. Such a function $R(s)=(\xi/L)^d
R_{W-D}[(\xi/L)^d s]$, which obviously obeys the sum rule (\ref{SR}),
 is not universal at {\it all} scales and reduces
to a constant $-(\xi/L)^d$ for $s\ll (L/\xi)^d$. Therefore,
in the limit (\ref{Lm})  the regular
part of $R(s)$ makes no contribution
to the r.h.s. of Eq. (\ref{NE}). Then $d\langle (\delta
N)^{2}\rangle/d\langle N\rangle$ is
exactly equal to 1 due to the  singular $\delta(s)$-term in $R(s)$.

Now we turn to microscopic calculations.
In the metallic region, $R(\omega)$ is given  by the two-diffuson
diagram \cite{AS}
that is convenient to represent (see for detail Ref.
\cite {AKL:91}) as in Fig.\ 1(a), separating
the diffusion propagators  (wavy lines). Both in the
metallic region for $\omega\raisebox{-.4ex}{\rlap{$\sim$}}
\raisebox{.4ex}{$<$}\Delta$ and at the mobility edge one
should consider also $2n$-diffuson corrections (Fig.\ 2(a) for n=2).
In all
diagrams, the
polygons with $2n+1$ vertices  are made
from the electron Green's functions that decrease exponentially over the
distance of the mean free path. Thus, all vertices of any polygon
correspond to the same spatial coordinate and its
ensemble-averaged contribution reduces to a constant which we
denote $\nu_0\tau^{2n}\chi_{2n+1}$, where $\chi_{2n+1}$ are dimensionless
complex numbers. Then  the general expression for the $2n$-diffuson
diagram in the momentum representation is given by
\begin{eqnarray}
R_{2n}(\omega)=
\frac{\Delta^{2n}|\chi_{2n+1}|^{2}}{\pi^{2}\beta}{\rm Re}\!\!
\sum_{q_1\cdots q_{2n}}\!\!
\left\{\delta_{\bf q}
\prod_{j=1}^{2n}\!P(\omega,{\bf q}_j)\!\right\}.
 \label{2n}
 \end{eqnarray}
Here $ P(\omega,{\bf q})$ is the {\em exact} diffusion
propagator, $\delta_{\bf q}\equiv
\delta_{q_1+\cdots+q_{2n},0}$, and
the factor $\beta^{-1}$  accounts for the number of diagrams in
different ensembles where some channels of propagation are suppressed.

In the metallic region, $P(\omega,{\bf q})=(Dq^2-i\omega)^{-1}$.
For $\omega\tau_D\!\ll\!  1$ (that corresponds to $t\gg\tau_{D}$), the excess
particle density is distributed homogeneously over the whole sample so that
only $q=0$ contribution of each diffuson survives in Eq.\ (\ref{2n}).
For $\omega\gg\Delta$, only
 the two-diffuson diagram  ($n=1$) is relevant \cite{AS} so that (with
$\chi_{3}=i$)
\begin{equation}
R(s)=- {(\pi^{2}\beta s^2)^{-1}}.
\label{R2}
\end{equation}

  The sum rule (\ref{SR}) allows to
calculate $\langle(\delta N)^{2}  \rangle$ in the energy interval
where $\langle N\rangle\gg1$ using only
the perturbative result (\ref{R2}). One represents the first term in
Eq.\ (\ref{V}) as
$-2\left< N \right>\int_{\left< N \right>}^{\infty}R(s)\,ds$
which is
a constant of order $1$. The second term in Eq.\ (\ref{V})
diverges only logarithmically.
Restricting it to the perturbative region with  a cutoff at
$s\raisebox{-.4ex}{\rlap{$\sim$}} \raisebox{.4ex}{$>$} 1$, one reproduces
the Wigner-Dyson result \cite{Wig}
with the accuracy up to a constant of order $1$:
\begin{equation}
\langle(\delta N)^{2} \rangle=\frac{2}{\pi^{2}\beta}\,\ln\langle N\rangle.
\label{WD}
\end{equation}

For   $E\gg 1/\tau_D$, this result does not hold
  in the metal \cite{AS} where such a width is unreachable
in the universality limit (\ref{Lm}), though.
On the contrary, at the mobility edge
$\Delta\sim 1/\tau_D$ and any interval with $\langle N\rangle\gg 1$
has the width
$E\gg 1/\tau_D$.  That is why one expects the variance
$\langle(\delta N)^{2} \rangle$ to deviate drastically from that in
Eq.\ (\ref{WD}).

 At the mobility
edge, $P(\omega,{\bf q})$ may be expressed as

 \begin{eqnarray}
 \label{DP}
 P(\omega,{\bf q})=\left[  D(\omega,{\bf q})q^2-i\omega  \right]^{-1}.
 \end{eqnarray}
which is the most general expression
compatible with the particle conservation law.
Although the exact diffusion coefficient $D(\omega,{\bf q})$
here is unknown,
the scaling and analytical properties of the diffusion propagator enable us to
determine $R(s)$ for $s\gg 1$.
Since the propagator $\tilde{P}(t,{\bf r}\!-\!{\bf r}')$, that is
the space-time Fourier-transform of $P(\omega,{\bf q})$,
is nonzero only for $t>0$ (causality) and real, $P(\omega,{\bf q})$
is analytical in the upper half-plane of the complex variable
$\omega$ and satisfies the relation
$P^{*}(\omega, {\bf q})=P(-\omega,-{\bf q})$. Using also the spatial isotropy,
one has $P^{*}(\omega, q)=P(-\omega, q)$.

At the mobility edge   in the limit $L\rightarrow\infty$,
the scaling arguments allow to express $P (\omega, q)$
in terms of the
dimensionless scaling function $F$ depending on
 $qL_{\omega}$, the ratio of the
only two lengths
characterizing the system\cite{JC1}. Here $ L_\omega $ is a characteristic
length of the displacement of a diffusing particle
for the time $\omega^{-1}$. At the critical point, a dimensional estimation
yields
\begin{equation}
 L_\omega \sim (\omega\nu_{0})^{-1/d}.
\label{Lo}
\end{equation}
With the
standard definition $L_{\omega}=|D(\omega)/\omega|^{1/2}$,
  Eq.\ (\ref{Lo}) reproduces
the well-known scaling result \cite{Weg:76}\nocite{ShaAb,Im:82}
\begin{equation}
\label{Do}
D(\omega)\propto  L_\omega ^{2-d}\propto\omega^{1-2/d}.
\end{equation}
Using the scaling relation (\ref{Lo}), we obtain
\begin{equation}
\label{GF}
P(\omega,q)=
(-i\omega)^{-1}
F\left(z \right),\qquad
z\equiv -i\omega\nu_{0}q^{-d},
 \end{equation}
where due to the above analyticity requirements,
$\omega$ contains an infinitesimal imaginary part, and
the function $F(z)$ is analytical for Re$z>0$ and
satisfies the condition
\begin{equation}
F^{*}(z)=F(z^{*}).
\label{An}
\end{equation}
In the static limit $P(\omega\rightarrow 0,q)\propto q^{-d}$
at the critical point\cite{Weg:80a}. In the opposite limit,
$ L_\omega q\ll1$, the diffusion propagator has the
form (\ref{DP}) with the
diffusion coefficient (\ref{Do}) depending only on $\omega$.
That results in the asymptotics
 \begin{eqnarray}
\label{z}
 F(z)\!=\! \left\{
 \begin{array}{llr}
 {\alpha_{1}\,z,}&{|z|\ll1} \\[2pt]
 {
 [{1+\alpha_{2}\,z^{-2/d}}] ^{-1}
\approx 1-
{\alpha_{2}}{z^{-2/d}} ,}&{|z|\gg1}
 \end{array}
 \right.
 \end{eqnarray}
 where $\alpha_{1,2}$ are real coefficients of order $1$.

Now we substitute Eq.\ (\ref{GF}) into Eq.\ (\ref{2n}), change $\sum_{q_j}
$for $L^d\int d^dq_j/(2\pi)^d$, and
represent $\delta_{\bf q}$ as
$\int d^dr\exp(i{\bf r}\sum {\bf q}_j)$ . Dividing the integration
over $\bf r$ into that over the
surface ($S_{d}$) and radius of the $d$-dimensional sphere, and
introducing dimensionless
variables $k_{j}=q_{j}r$ and $\zeta=\omega\nu_0r^{d}$,
we reduce Eq.\ (\ref{2n}) to
\begin{eqnarray}
R_{2n}&=&\frac{\Delta}{\omega}\frac{ (-1)^n
|\chi_{2n+1}|^{2}}{\beta\pi^{2}d}
   \!\int\! dS_{d}\prod_{j=1}^{2n}
\int\!\frac{d^{d}k_{j}}{(2\pi)^{d}}
\,e^{ik_{j}\cos\theta_{j}}\nonumber\\&\times&
\int_{0}^{\infty}
 \!{d\zeta}\,{\zeta^{-2n}}
{\rm Re}\prod_{j=1}^{2n}
F\left( {-i\zeta}{k_{j}^{-d}} \right).
\label{0}
\end{eqnarray}
A dimensional estimation of this integral would give ${\Delta}/{\omega} $
so that $R(s)\sim 1/s$. Having
substituted this into  Eq.\ (\ref{V}), one would obtain
$\langle(\delta N)^{2} \rangle\sim\left< N \right>\ln\left< N \right>$ which
is strictly prohibited by the sum rule, as shown above.

However, it
follows from Eq.\ (\ref{An}) that the real part of the product of
$F$-functions
in Eq.\ (\ref{0}) is an even function of $\zeta$. Thus, the
integration over $\zeta$ can be extended to the whole real axis.
Taking into account the asymptotics   (\ref{z})
and the analyticity of the function $F(-i\zeta)$
in the upper half-plane of the
complex variable $\zeta$,
one concludes immediately that the integral (\ref{0}) equals zero.

Therefore, $R(\omega)=0$ for $\omega\gg\Delta$ in the limit (\ref{Lm}).
For large but finite $L$, one
has to consider corrections to the diffusion propagator proportional to powers
of the small parameter
$ L_\omega /L=(\omega\nu_{0}L^{d})^{-1/d}=
(\Delta/\omega)^{1/d}$:
\begin{equation}
P(\omega,q)=(-i\omega)^{-1}\left[
F\left(z\right)+
\left( {\Delta}/{-i\omega} \right)^{1-\gamma}
\Phi(z)\right],
\label{Fi}
\end{equation}
where the scaling function $\Phi(z)$ has the same analytical
properties as $F(z)$.

To find $\gamma$ one uses Eq.\ (\ref{DP}) in the limit
 $ L_\omega q\ll1$. Substituting there $D(\omega)\propto L_{\omega}^{2-d}
\left[1+(L_{\omega}/L)^{1/\nu}\right]$ (resulting from the standard
renormalization group equation)
instead of Eq.\ (\ref{Do}), one expands the diffusion propagator
up to the first power in $D(\omega)q^{2}/\omega$. Comparing such an expansion
to Eq.\ (\ref{Fi}), we have:
\begin{equation}
\gamma=1-(\nu d)^{-1}.
\label{G}
\end{equation}
Note that $1/2\!<\!\gamma\!<\!1$ due to the Harris criterion\cite{Har}.

Repeating the procedure which led to Eq.\ (\ref{0}) with
$P(\omega,q)$ given by Eq.\ (\ref{Fi}), we obtain:
\begin{eqnarray}
\lefteqn{R_{2n}(\omega)=\frac{n(-1)^{n}S_{d}|\chi_{2n+1}|^{2}}{\beta\pi^{2}d}
\frac{\Delta}{\omega}  \prod_{j=1}^{2n}
\int\frac{d^{d}k_{j}}{(2\pi)^{d}}
\,e^{ik_{j}\cos\theta_{j}} }\nonumber\\
&&\times\int_{0}^{\infty}\!\!\frac{d\zeta}{\zeta^{2n}}
{\rm Re}\left\{\!\left(\frac{\Delta}{-i\omega} \right)^{\! 1-{\gamma}}\!\!
\Phi\!\left( \frac{-i\zeta}{k_{2n}^{d}} \right)\!
\prod_{j=1}^{2n-1}\!F\biggl( \frac{-i\zeta}{k_{j}^{d}} \biggr) \!\right\}\!.
\label{I}
\end{eqnarray}
Here, in contrast to Eq.\ (\ref{0}), the integrand has
an odd in $z$ part.
This is the only part which contributes to the integral (\ref{I}).
As this integral is a nonzero dimensionless number,
we obtain using Eq.\ (\ref{G}):
\begin{equation}
\label{Ro}
R(s)=-{c_{d\beta}}\beta^{-1}s^{-2+\gamma}
\qquad(s\equiv\omega/\Delta \gg 1),
\end{equation}
where $c_{d\beta}$ is a numerical factor. For $\beta =1$,
 $d=2+\epsilon$
expansion gives $\nu=1/\epsilon$ and $\gamma=2/d$ near $d=2$.
In this case, the integrand in Eq.\ (\ref{I}) has no odd part  and
$c_d$ vanishes at $d=2$.

With $R(s)$ from Eq.\ (\ref{Ro})  the integral in the sum rule
(\ref{SR}) is convergent, and we can use it
for calculating the first integral in Eq.\ (\ref{V}).
The second integral in Eq.\ (\ref{V}) is also determined by the region
$s\sim \left< N \right>\gg 1$,
and we arrive at the announced result (\ref{1}), where
$a_{d}=2c_{d}/\gamma(1-\gamma)$.

Since the coefficient $a_{d}$ must be positive,   $c_{d}>0$,
and the correlator $R(\omega)$ is negative for $\omega\gg\Delta$.
For small $\omega\!\ll\! \Delta$ one can use the same zero-mode approximation
\cite{Ef:83}
as in the metal region for $\omega\!\ll\! 1/\tau_{D}$, so
that the correlation function
$R(s)$ should have the Wigner-Dyson form. We can conclude, therefore, that the
energy levels are repelling at all energy scales.

 Note in conclusion that the Wigner-Dyson statistics
  can be represented
as the Gibbs statistics of a classical one-dimensional gas of fictitious
particle with the pairwise interaction $V(s\!-\!s')=-\!\ln|s\!-\!s'|$. The
Poisson
statistics corresponds to $V(s\!-\!s')=0$. If we suppose that
the statistics of energy levels in the critical
region   can  also be represented as a Gibbs
statistics with some pairwise interaction $V(s\!-\!s')$, then such an
effective interaction may be found,
using the approach  developed  for the random matrix theory
\cite{Wig,RMT,Bee}.  Thus, in order to
reproduce the asymptotics of the two-level
correlation function (\ref{Ro}),
 the interaction should have the form \cite{KLA}:
 \begin{eqnarray}
 \label{In}
 V(s\!-\!s')=\frac{1-\gamma}{2\pi c_{d\beta}}\cot(\pi\gamma/2)
 \frac{1}{|s-s'|^{\gamma}}.
 \end{eqnarray}
This interaction is valid for $s\!-\!s'\gg 1$. For small \mbox{$s-s'$} the
interaction should be of the Wigner-Dyson form. Therefore, $V(s-s')$  always
remains repulsive.

In order to check the conjecture about a pairwise nature of the
effective interaction, one should investigate the higher order
correlation functions.
If they are factorizable
 like in the random matrix theory
\cite{RMT}, then the Gibbs model with the interaction (\ref{In}) will
describe the whole statistics at the mobility edge.

We are thankful to the ICTP at Trieste for  giving us opportunity to
meet together. V.E.K.\ thanks the University of Birmingham for
kind hospitality extended to him during the
initial stage of this work.
V.E.K.\ and I.V.L.\ acknowledge travelling support  under the
EEC contract No.\ SSC-CT90-0020. Work at MIT was
 supported by the NSF under Grant No. DMR92-04480.


\end{document}